\documentclass[floatfix,twocolumn,aps,prb,showpacs,amsmath,amssymb]{revtex4-1}
\usepackage[latin1]{inputenc}
\usepackage[dvips]{graphicx}

\usepackage{dcolumn}
\usepackage{bm}
\usepackage{dsfont}
\usepackage{color}
\usepackage{url}
\usepackage[colorlinks=true ,citecolor=blue]{hyperref}
\usepackage{amsmath,amssymb,esint}
 \usepackage[table]{xcolor}

\definecolor{Nathanblue}{rgb}{0.,0.24,0.51}

\def\bs#1{\boldsymbol{#1}}

\newcommand{\be}{\begin{equation}}
	\newcommand{\ee}{\end{equation}}
\newcommand{\bq}{\begin{eqnarray}}
	\newcommand{\eq}{\end{eqnarray}}

\begin{document}

\title{Tensor Berry connections and their topological invariants}

\author{Giandomenico Palumbo and Nathan Goldman}
\affiliation{Center for Nonlinear Phenomena and Complex Systems,
	Universit\'e Libre de Bruxelles, CP 231, Campus Plaine, B-1050 Brussels, Belgium}

\date{\today}

\begin{abstract}

The Berry connection plays a central role in our description of the geometric phase and topological phenomena. In condensed matter, it describes the parallel transport of Bloch states and acts as an effective ``electromagnetic" vector potential defined in momentum space. Inspired by developments in mathematical physics, where higher-form (Kalb-Ramond) gauge fields were introduced, we hereby explore the existence of ``tensor Berry connections" in quantum matter. Our approach consists in a general construction of effective gauge fields, which we ultimately relate to the components of Bloch states. We apply this formalism to various models of topological matter, and we investigate the topological invariants that result from generalized Berry connections. For instance, we introduce the 2D Zak phase of a tensor Berry connection, which we then relate to the more conventional first Chern number; we also reinterpret the winding number characterizing 3D topological insulators to a Dixmier-Douady invariant, which is associated with the curvature of a tensor connection. Besides, our approach identifies the Berry connection of tensor monopoles, which are found in 4D Weyl-type systems [Palumbo and Goldman, Phys.~Rev.~Lett.~121,~170401~(2018)]. Our work sheds light on the emergence of gauge fields in condensed-matter physics, with direct consequences on the search for novel topological states in solid-state and quantum-engineered systems.

\end{abstract}

\maketitle

\section{Introduction}

The geometric phase is recognized as one of the most important concepts in quantum mechanics~\cite{Pancharantnam,Higgins,AB,WuYang,Mead1,Berry,Simon,Simon2,Mead,Xiao,Dalibard,Goldman12,Goldman13}. It has numerous applications in diverse physical situations, ranging from electromagnetism and high-energy physics to atomic-molecular-optics and condensed-matter physics. In the realm of quantum matter, the ``Berry" phase plays an important role in the interpretation and exploration of topological states of matter~\cite{TKNN,Simon,Xiao,Fradkin}. In particular, the gauge field associated with the Berry phase, the so-called Berry connection~\cite{Xiao}, constitutes a central quantity in the definition of topological invariants characterizing topological matter, examples of which include the Zak phase~\cite{Zak} in one spatial dimension, the Chern numbers in two~\cite{TKNN} and four~\cite{Zhang4D} dimensions, and the Chern-Simons invariant~\cite{Zhang} in three dimensions; see also Refs.~\cite{Qi,Ryuclass,Ryu} on topological invariants and the classification of topological matter. 

The Berry connection transforms as a vector gauge potential under a gauge transformation~\cite{Xiao}. Hence, in lattice systems, the Berry connection that is associated with a given Bloch state acts as an effective ``electromagnetic" vector potential defined in quasi-momentum space~\cite{Xiao}. Gauge invariant quantities can be derived from the Berry connection, such as the Berry curvature, which acts as an effective ``electromagnetic" (Faraday) tensor in momentum space. This Berry-curvature field~\cite{Xiao}, which naturally enters the definition of the Berry phase and the aforementioned Chern numbers~\cite{Qi,TKNN}, was recently measured in cold atoms~\cite{Duca,Flaschner} and photonics~\cite{Price_Berry_exp}. The notions of geometric phase, Berry connection and curvature can be generalized to the case of degenerate spectra, where these gauge structures become non-Abelian~\cite{Wilczek,Pachos}. Altogether, gauge theories appear as a powerful and universal language to describe and explore emergent phenomena in quantum matter.

In gauge theories, various types of gauge fields exist beyond the well-known vector potential of electromagnetism. While different forms of \emph{vector} gauge fields have been introduced in the literature, such as the non-Abelian gauge potential of Yang-Mills theory~\cite{Yang-Mills}, recent theoretical developments have considered gauge fields of even more exotic nature, such as \emph{tensor ($p$-form) gauge fields}~\cite{Henneaux}. Tensor gauge fields, example of which include the so-called Kalb-Ramond fields~\cite{Kalb-Ramond}, play an important role in string theory as well as in (3+1)-dimensional topological field theories, the commonly-called BF theories~\cite{Thompson,Cattaneo}. It was recently shown in Refs.~\cite{Hansson, Moore, Fradkin2, Trugenberger, Palumbo} that BF theories can describe the real-space behavior of topological insulators and superconductors, in the low-energy regime. Yet, the relation between the emergent Kalb-Ramond fields and the physical variables of the underlying microscopic models remains to be explored.

In this work, we provide a general approach to effective gauge fields in quantum matter, which includes the Berry connection as a special example. In this framework, gauge fields are built from pseudo-real and complex scalar gauge fields, which are all ultimately related to the components of Bloch states of interest. This leads to a wide variety of ``Berry" connections of different nature, including ``flat" connections~\cite{Carpentier} and tensor gauge fields~\cite{Palumbo-Goldman}. We illustrate our formalism by considering emblematic microscopic models of topological matter, which allows us to reinterpret well-known topological invariants in terms of novel types of gauge fields, such as tensor Berry connections. As a first result of our approach, we will show how the topology of 1D topological insulators in class AIII~\cite{Ryu,Hughes,Cheng} can be evaluated through a Zak phase~\cite{Xiao} associated with an abstract connection~\cite{Carpentier}. Then, we will use the existence of tensor Berry connections to introduce higher-dimensional generalizations of the Zak phase~\cite{Xiao}; in particular, we will show how the first Chern number characterizing 2D Chern insulators can be viewed as the 2D Zak phase of a tensor connection. We will also analyze the curvature associated with tensor Berry connections, as well as its related topological invariant:~the so-called Dixmier-Douady ($\mathcal{DD}$) invariant~\cite{Murray1,Murray2}. As an intriguing result, we will demonstrate that the topological winding number~\cite{Ryu,Neupert} characterizing 3D topological insulators in class AIII is strictly equivalent to the $\mathcal{DD}$-invariant associated with a tensor Berry connection. Moreover, we will relate a 4D tensor monopole~\cite{Nepomechie,Teitelboim}, previously introduced in Ref.~\cite{Palumbo-Goldman}, to the tensor Berry connection of a 4D Weyl-type Hamiltonian. We will discuss how the existence of these exotic topological defects could give rise to novel forms of topological states in higher-dimensional systems. Finally, we will briefly mention possible generalizations of our findings to higher-order tensor connections.

This work is organized as follows. Section~\ref{section_scalar_vector} introduces our general approach, and it applies it to the case of scalar and vector gauge fields in quantum matter. The following Section~\ref{section_tensor}, which constitutes the core of the article, introduces the notion of tensor Berry connection; this Section illustrates this concept on a series of topological models. An overview of the topological structures that emanate from this work is summarized in Table~\ref{Table_one}, which is described in Section~\ref{Section_table}. Concluding remarks are provided in Section~\ref{Section_conclusions}.

This work sheds light on the emergence of gauge structures and geometric phases in condensed-matter physics, with direct consequences on the search for novel topological states in solid-state \cite{Qi} and quantum-engineered setups~\cite{Goldman_review,Cooper_review,Ozawa_review}.

%
%


\section{The construction applied to scalar and vector gauge fields}\label{section_scalar_vector}

As a warm-up, we first illustrate our approach by constructing scalar and vector gauge fields from the components of a general Bloch state $|u_n(\textbf{q})\rangle$, where $\textbf{q}$ refers to the quasi-momentum of a generic quantum lattice system; the band index $n$ will be omitted in the following. While the notion of scalar gauge field naturally arises from Bloch states, we will see that two types of vector gauge fields (connections) can be defined:~the conventional Berry connection~\cite{Xiao}, and a ``flat" connection~\cite{Carpentier} (whose related curvature is trivial). We will show that this novel approach to Berry connections provides an instructive interpretation of the one-dimensional (1D) winding number of Refs.~\cite{Ryu,Hughes} in terms of a Zak phase associated with a peculiar Berry connection. This formal construction of Berry connections will then be generalized in Section~\ref{section_tensor}, in view of introducing the concept of tensor Berry connection.

We point out that the construction and developments presented in this work can be applied to general quantum systems described by a Hamiltonian $\hat H (\bs{p})$, where $\bs{p}\!\in\!\mathcal{M}$ are parameters defined over some parameter space $\mathcal{M}$. For the sake of concreteness, we have decided to present our approach using the framework offered by Bloch states in lattice systems, in which case the parameters $\bs{p}\!\rightarrow\!\textbf{q}$ correspond to the quasi-momenta and the relevant parameter space $\mathcal{M}$ corresponds to the first Brillouin zone (FBZ).

\subsection{Scalar gauge fields from Bloch states}

In quantum field theory, a complex scalar (matter) field undergoes the standard U(1) gauge transformation
\begin{eqnarray}
\phi\rightarrow e^{i \alpha} \phi, \qquad A_{\mu}\rightarrow A_{\mu}+\partial_{\mu}\alpha ,\label{gauge_transf}
\end{eqnarray}
where the function $\alpha$ depends on the coordinates; here $A_{\mu}$ denotes the $\mu$-th component of a vector gauge field that is coupled to the matter field $\phi$. A simple instance of such gauge structures is found in the Abelian Higgs model, where a complex field $\phi$ is coupled to an Abelian gauge field $A_{\mu}$. Following the approaches of Refs.~\cite{Dvali,Scalar}, one can also introduce a real scalar field $\varphi$, which transforms as
\begin{eqnarray}\label{scalar-field}
\varphi\rightarrow \varphi+\alpha ,
\end{eqnarray}
under the U(1) gauge transformation in Eq.~\eqref{gauge_transf}. Such scalar gauge fields play an important role in the context of axion electrodynamics~\cite{Wilczek2,Dvali,Scalar,Axions}.\\

Now, consider the problem of a quantum particle moving on a lattice. The corresponding spectrum displays a band structure, $E_n (\textbf{q})$, and the associated eigenstates are the Bloch states $\vert u_n(\textbf{q})\rangle$. In the following, we will consider a given (non-degenerate) Bloch state, and we will write its components as $\vert u \left (\textbf{q})\rangle\!=\!(u^{1}(\textbf{q}),u^{2}(\textbf{q}),...,u^{N}(\textbf{q}) \right )^{\top}$, where $N$ will be assumed to be finite (i.e.~the examples discussed below are based on $N$-band models); each component with be denoted $u^{\aleph}(\textbf{q})$, with $\aleph\!=\!1,2,\dots,N$.

Such a non-degenerate Bloch state has a gauge degree of freedom, in the sense that it undergoes the transformation
\begin{eqnarray}\label{Bloch}
|u ({\bf q}) \rangle\rightarrow e^{i\alpha({\bf q})} |u ({\bf q}) \rangle,
\end{eqnarray}
under a local gauge transformation~\cite{Xiao}; this gauge degree of freedom is consistent with the fact that a wave function does not constitute a physical observable. From Eq.~\eqref{Bloch}, one identifies two families of scalar fields that are defined in $\textbf{q}$-space: 
\begin{itemize}
\item each component $u^{\aleph}(\textbf{q})$ of the Bloch state, defines a \emph{complex scalar field}; 
\item a \emph{pseudo-real scalar gauge field} can be constructed from the Bloch state as
\begin{eqnarray}\label{scalar}
\varphi(\textbf{q})=-\frac{i}{N}\log \prod\limits_{\aleph} u^{\aleph}(\textbf{q}).
\end{eqnarray}
\end{itemize}
Indeed, it is straightforward to show that the complex fields $u^{\aleph}(\textbf{q})$ transform as in Eq.~\eqref{gauge_transf}, and that the pseudo-real scalar gauge field $\varphi(\textbf{q})$ in Eq.~\eqref{scalar} satisfies the gauge transformation defined in Eq.~\eqref{scalar-field}.
Here, the wording \emph{pseudo-real scalar} refers to the fact that $\varphi(\textbf{q})$ can a priori be a complex-valued function (which transforms according to Eq.~\eqref{scalar-field}).
These notions of complex and pseudo-real scalar gauge fields built from Bloch states are central in our construction of tensor Berry connections [Section~\ref{section_tensor}].

\subsection{Vector gauge fields from Bloch states}\label{section_vector}

\subsubsection{Vector gauge fields, geometry and topology: \\ A brief reminder}

Vector gauge fields are central in gauge theories, in particular, in the context of high-energy physics. In general, an Abelian vector gauge field (or connection) $A_{\mu}$ transforms as 
\begin{eqnarray}
A_{\mu}\rightarrow A_{\mu}+\Lambda_{\mu},
\end{eqnarray}
under a $U(1)$ gauge transformation. Here, $\Lambda_{\mu}$ is an arbitrary (1-form) field, which depends on the coordinates. In specific cases, $\Lambda_{\mu}$ can itself be expressed as an exterior derivative, such that
\begin{eqnarray}\label{A-field}
A_{\mu}\rightarrow A_{\mu}+\partial_{\mu}\alpha.
\end{eqnarray}
This is the case for the conventional vector potential of electromagnetism. 

From the vector gauge field, $A_{\mu}$, one can define the so-called Wilson loop, 
\begin{eqnarray}
W_{L}=\exp \left( i  \int_{\gamma} \text{d} x^{\mu}\, A_{\mu}\right),\label{Wilson}
\end{eqnarray}
which involves the circulation of the gauge field around a closed path $\gamma$ in coordinate space (space-time). Mathematically, this geometric quantity is associated with the mismatch of parallel transport upon performing the loop $\gamma$, a concept known as holonomy in fibre bundle theory~\cite{WuYang,Simon,Nakahara}. In electromagnetism, this Wilson loop corresponds to the Aharonov-Bohm phase factor acquired by a wavefunction in the presence of a finite magnetic flux~\cite{WuYang}. Using Stokes' theorem, the line integral in Eq.~\eqref{Wilson} can be re-expressed as the surface integral of the curvature $\mathcal{F}\!=\!\text{d} A$, whose components are given by
\begin{eqnarray}\label{curvatureF}
\mathcal{F}_{\mu\nu}=\partial_{\mu}A_{\nu}-\partial_{\nu}A_{\mu}.
\end{eqnarray}
In the case of electromagnetism, the curvature corresponds to the Faraday (field-strength) tensor. In general, the curvature $\mathcal{F}_{\mu\nu}$, and hence, the Wilson loop, represent gauge invariant quantities associated with the Abelian gauge field $A_{\mu}$.

Beyond the geometry captured by the Wilson loop, the curvature $\mathcal{F}$ is also deeply connected to topology, in particular, to the so-called Chern classes~\cite{Nakahara}. When integrating $\mathcal{F}$ over a compact space-time manifold $\mathcal{M}$ of dimension two (e.g.~a sphere $S^2$), one obtains a topological invariant (an integer) known as the first Chern number
\begin{eqnarray}
\nu^{1}=\frac{1}{2 \pi} \int_{\mathcal{M}} \text{d}^{2}x\, \mathcal{F}_{xy} , \qquad \text{dim} \mathcal{M}=2.\label{Chern_def}
\end{eqnarray}
In electromagnetism, this integer corresponds to the magnetic charge of a monopole~\cite{WuYang,Nakahara} located at the center of the sphere $S^2$. Another topological invariant originating from the Chern classes, the so-called second Chern number~\cite{Simon2,Zhang4D,Goldman3}, can be obtained by integrating $\mathcal{F}^{2}\equiv\mathcal{F}\wedge \mathcal{F}$ over a manifold $\mathcal{M}$ of dimension four (e.g.~a sphere $S^4$),
\begin{equation}
\nu^{2}\!=\!\frac{1}{4 \pi^{2}}\int_{\mathcal{M}} \text{d}^{4}x\, (\mathcal{F}_{xy}\mathcal{F}_{zw}+\mathcal{F}_{zy}\mathcal{F}_{wx}+\mathcal{F}_{yw}\mathcal{F}_{zx}). \label{Chern2_def}
\end{equation}
Formally, the topological (and integral) nature of the first and second Chern numbers can be understood in terms of the cohomology groups $H^{2}(S^{2})\!=\!\mathbb{Z}$ and $H^{4}(S^{4})\!=\!\mathbb{Z}$, respectively~\cite{Nakahara}. Higher-order Chern numbers, $\nu^{n}$, can be defined by considering higher powers of the curvature, $\mathcal{F}^n$, integrated over higher-dimensional spaces~\cite{CHLee6D,Price6D}.


\subsubsection{Constructing vector connections from Bloch states}\label{construction_vector}

The gauge structure associated with the Bloch state $|u ({\bf q}) \rangle$, Eq.~\eqref{Bloch}, naturally gives rise to vector gauge potentials defined in $\bf{q}$-space, an example of which is the well-known Berry connection~\cite{Xiao}. In this Section, we analyse the existence of such ``momentum-space" vector potentials using a formal approach. This will set a framework for the introduction of ``tensor Berry connections" in Section~\ref{section_tensor}.

We start by introducing two sets of scalar fields, 
\begin{equation}
\phi_{1}^{\aleph}\!=\!\phi^{\aleph}_{1}( |u ( {\bf q}) \rangle ), \quad \phi_{2}^{\aleph}\!=\!\phi^{\aleph}_{2}( |u ( {\bf q}) \rangle ) ,
\end{equation}
where the index $\aleph\!=\!1,2,\dots,N$ is used to label the different fields. At this stage, we assume that these scalar fields depend on the various components $u^{\aleph}(\textbf{q})$ of the Bloch state $|u ({\bf q}) \rangle$; hence, these fields implicitly depend on the quasi-momenta $\textbf{q}$. Moreover, we will take these scalar fields to be complex, and we set their gauge transformations to be of the form
\begin{align}\label{gauge}
&\phi_{1}^{\aleph}\rightarrow e^{-i \alpha(\textbf{q})} \phi_{1}^{\aleph}, \\
&\phi_{2}^{\aleph}\rightarrow e^{i \alpha(\textbf{q})} \phi_{2}^{\aleph},\notag
\end{align}
under any local $U(1)$ gauge transformation [Eq.~\eqref{Bloch}]. 

Next, we define the following gauge connection
\begin{eqnarray}\label{vectorB}
A_{\mu}=\frac{i}{2}\,\epsilon^{jk} \phi_{j}^{\aleph}\partial_{\mu}\phi_{k}^{\aleph}, \quad j,k=\{1,2\},
\end{eqnarray}
where $\partial_{\mu}\equiv \partial_{q_{\mu}}$, and where an implicit summation over the field index $\aleph$ is assumed. By substituting Eq.~\eqref{gauge} into Eq.~\eqref{vectorB}, we obtain a transformation law for the gauge connection,
\begin{eqnarray}
A_{\mu}= \frac{i}{2}\,\phi_{1}^{\aleph}\partial_{\mu}\phi_{2}^{\aleph}- \frac{i}{2}\,\phi_{2}^{\aleph}\partial_{\mu}\phi_{1}^{\aleph}-(\phi_{1}^{\aleph}\phi_{2}^{\aleph})\partial_{\mu}\alpha.
\end{eqnarray}
This corresponds to the gauge transformation of a vector gauge potential [Eq.~\eqref{A-field}], if and only if
\begin{eqnarray}\label{Cons}
\phi_{1}^{\aleph}\phi_{2}^{\aleph}=1.
\end{eqnarray}
This condition, when combined with Eq.~\eqref{gauge}, is satisfied in two different cases, as we know discuss. 

\paragraph{The Berry connection.} The conditions in Eqs.\eqref{gauge} and \eqref{Cons} are satisfied by imposing the constraints
\begin{equation}
	\phi_{1}^{\aleph}=(\phi_{2}^{\aleph})^{*}, \quad \sum_{\aleph} \vert \phi_{2}^{\aleph} \vert^2 = 1.
\end{equation}
This is naturally satisfied by choosing an explicit relation between the scalar fields and the components of the Bloch state, $\phi_{2}^{\aleph}\!\equiv\!u^{\aleph}(\textbf{q})$. Doing so, the connection in Eq~\eqref{vectorB} acquires the form of the well-known (Abelian) Berry connection,
\begin{equation}
	A_{\mu}=i\langle u|\partial_{\mu}|u\rangle ,\label{Berry_def}
\end{equation}
which indeed acts as a vector potential in ${\bf q}$-space~\cite{Xiao}. 

The geometric and topological structures associated with the Berry connection are well established~\cite{Xiao}, as we now briefly review. The analogue of the Wilson loop in Eq.~\eqref{Wilson}, which involves the line integral of the Berry connection over a closed path in ${\bf q}$-space, corresponds to the famous geometric (Berry) phase~\cite{Berry}. When performing this line integral over the entire first Brillouin zone (FBZ) of a 1D lattice system,
\begin{eqnarray}
	\gamma^{\text{1D}}=\int_{\text{FBZ}} A (q) \text{d} q,\label{Zak_def}
\end{eqnarray}
one obtains the so-called Zak phase~\cite{Zak}, which is a gauge invariant quantity that plays a central role in the theory of polarization~\cite{Xiao,Vanderbilt,Resta}. Furthermore, when the underlying Hamiltonian exhibits a discrete symmetry (e.g.~inversion, particle-hole), the Zak phase $\gamma^{\text{1D}}$ is quantized and defines a topological invariant characterizing the related symmetry-protected phases~\cite{Hatsugai1,Hatsugai2}. Finally, the curvature [Eq.~\eqref{curvatureF}] associated with the Berry connection, $\mathcal{F}\!=\!\text{d} A$, the so-called Berry curvature~\cite{Xiao}, acts as a field-strength tensor in ${\bf q}$-space. Integrating the Berry curvature over the entire FBZ of a 2D system leads to the first Chern number [Eq.~\eqref{Chern_def}], which is used to characterize Chern insulators~\cite{TKNN}; similarly, a second Chern number [Eq.~\eqref{Chern2_def}] can be introduced to characterize topological insulators in 4D lattice systems~\cite{Zhang4D,Goldman3}. We remind the reader that a Berry curvature leading to a non-zero Chern number takes the form a monopole field in ${\bf q}$-space; such fictitious monopoles, defined in the parameter-space of a quantum system, were recently explored in ultracold atoms~\cite{Ray,Spielman}.

\paragraph{The flat connection.} The conditions in Eqs.\eqref{gauge} and \eqref{Cons} can also be satisfied by the alternative constraint
\begin{eqnarray}
\phi_{1}^{\aleph}=(1/N) \left (\phi_{2}^{\aleph} \right )^{-1}.\label{flat_constraint}
\end{eqnarray}
Setting $\phi_{2}^{\aleph}\equiv u^{\aleph}(\bs q)$ as above, the corresponding connection [Eq~\eqref{vectorB}] then takes the form
\begin{align}\label{scalar2}
\tilde A_{\mu}&=\frac{i}{2N}\,\frac{1}{\phi_{2}^{\aleph}}\partial_{\mu}\phi_{2}^{\aleph}-\frac{i}{2N}\,\phi_{2}^{\aleph}\,\partial_{\mu}\frac{1}{\phi_{2}^{\aleph}} \nonumber \\ 
&=\frac{i}{N} \frac{1}{\phi_{2}^{\aleph}} \partial_{\mu}\phi_{2}^{\aleph}=\frac{i}{N}\,\partial_{\mu}\log \Pi_{\aleph}\phi_{2}^{\aleph}=-\partial_{\mu}\varphi,
\end{align}
where the last step involves the scalar gauge field $\varphi$ previously defined in Eq.~\eqref{scalar}. This last result shows that the curvature of this alternative connection, $\tilde{\mathcal{F}}\!=\!\text{d} \tilde{A}=0$, is trivial. In other words, the connection~\eqref{scalar2}, which stems from the contraint in Eq.~\eqref{flat_constraint}, is ``flat"; see also Ref.~\cite{Carpentier}. 

\paragraph{Non-trivial topology of the pseudo-real scalar field.}

While being flat in two and higher dimensions, the Abelian connection in Eq.~\eqref{scalar2} can nevertheless be used to characterize the topological nature of 1D topological insulators in class AIII~\cite{Ryu,Hughes,Cheng}, as we now demonstrate. 

Let us consider the following Hamiltonian~\cite{Cheng}
\begin{equation}\label{1D}
H_{\text{1D}}=
S_{1}\sigma_{x}+S_{2}\sigma_{y} = \sin q \sigma_{x} + \left ( M-\cos q \right )\sigma_{y} ,
\end{equation}
where $\sigma_{x,y}$ are Pauli matrices, and where $q$ denotes the quasi-momentum in a 1D lattice system.
The eigenstates corresponding to the lower band are then given by
\begin{equation}
\vert u_{-} \rangle=\frac{1}{\sqrt{2}}(u_{1},1)^{\top}, \, \quad u_{1}=\frac{S_{1}-i S_{2}}{|E|},\label{u_minus}
\end{equation}
with their energy given by $E\!=\!-\sqrt{S_{1}^{2}+S_{2}^{2}}$. When this lower band is completely filled, the system is in an insulating state, and its topological class is established by the topologically-invariant winding number~\cite{Ryu,Hughes},
\begin{align}
w^{1}=\frac{1}{2\pi}\int_{\text{FBZ}}\text{d}q\, \epsilon^{jk}\frac{1}{|E|^{2}} S_{j} \left ( \partial_{q}S_{k} \right ).\label{winding_def}
\end{align}
For the Hamiltonian defined in Eq.~\eqref{1D}, this winding number is $w^{1}\!=\!1$ whenever $-1\!<\!M\!<\!1$, and zero otherwise (the bulk gap closes at $|M|\!=\!1$).

Interestingly, the winding number in Eq.~\eqref{winding_def} can be interpreted as the integral of a connection $\tilde A_{\mu}\!=\!-\partial_{q}\varphi$ generated by a well-defined pseudo-scalar field $\varphi$, in analogy with the Zak phase in Eq.~\eqref{Zak_def}. Indeed, considering the pseudo-scalar gauge field built from the lower-band eigenstates [Eq.~\eqref{scalar}],
\begin{equation}
\varphi_-(q)=-\frac{i}{2}\log \prod\limits_{\aleph=1,2} u^{\aleph}_-(q),\label{scalar_1D}
\end{equation}
one obtains that the Zak phase associated with the connection $\tilde A_-(q)\!=\!-\partial_{q}\varphi_-$ yields [Eqs.~\eqref{u_minus}-\eqref{scalar_1D}]
\begin{align}
\int_{\text{FBZ}} \tilde A_- (q) \, \text{d}q = -\int_{\text{FBZ}}\text{d}q\,\partial_{q}\varphi_-= \pi w_{1},\label{Zak_flat}
\end{align}
which, as we announced, is directly proportional to the winding number in Eq.~\eqref{winding_def}.

As already pointed out in Ref.~\cite{Carpentier}, the line-integral of a flat connection, around a generic closed path, can indeed be non-zero if this path is non-contractible to a point; this condition is naturally satisfied by the Zak phase in Eq.~\eqref{Zak_flat}, which involves a non-contractible loop around the FBZ of a 1D system.

This simple example illustrates how a topological invariant, used to classify topological states of matter, can be re-interpreted in terms of the Zak phase of an unusual connection [Eq.~\eqref{scalar2}], which differs from the more conventional ``Berry connection" in Eq.~\eqref{Berry_def}. This idea will be further exploited in the next Section, which concerns tensor generalizations of the Berry connection.

\section{Tensor Berry connections}\label{section_tensor} 

In this Section, we investigate the existence of tensor connections in condensed matter, and we present a couple of applications in topological systems.

\subsection{Tensor gauge fields and topology}

Let us first briefly remind some general definitions and properties related to (2-form) tensor gauge fields. An Abelian 2-form gauge field, 
\begin{equation}
B\!=\!B_{\mu\nu} \text{d}x^{\mu}\wedge \text{d}x^{\nu}, \qquad B_{\nu\mu} = - B_{\mu\nu},\label{B_anti}
\end{equation}
is an object whose components are labelled by two-indices, and which satisfies the following gauge transformation~\cite{Dvali,Gaiotto,Montero} (or local shift symmetry),
\begin{eqnarray}\label{BB-field}
B_{\mu\nu}\rightarrow B_{\mu\nu}+\Lambda_{\mu\nu},
\end{eqnarray}
where $\Lambda_{\mu\nu}$ is a 2-form field; see Ref.~\cite{Nakahara} for details on differential forms, their wedge product ($\wedge$) and exterior derivative. In some cases, $\Lambda_{\mu\nu}$ can itself be expressed as an exterior derivative, $\Lambda_{\mu\nu}\!=\!\partial_{\mu}\xi_{\nu}\!-\!\partial_{\nu}\xi_{\mu}$, where $\xi_{\mu}$ is a 1-form.

Similarly to the Wilson loop of a vector gauge field [Eq.~\eqref{Wilson}], one can define a ``Wilson surface" associated with a tensor gauge field~\cite{Cattaneo}
\begin{equation}
	W_{S}\!=\!\exp\left( i \iint B \right)\!=\!\exp\left( i \iint dx^{\mu}\wedge dx^{\nu}\, B_{\mu\nu}\right). \label{Wilsonsurface}
\end{equation}
This represents a direct generalization of the Aharonov-Bohm (geometric) phase to 2-form gauge fields, which is related to the notion of surface (or gerbe) holonomy in differential geometry~\cite{Waldorf}. Using Stokes' theorem, the Wilson surface can be expressed in terms of a 3-form curvature, $\mathcal{H}\!=\!\text{d} B$, whose components are given by
\begin{eqnarray}
	\mathcal{H}_{\mu\nu\lambda}=\partial_{\mu}B_{\nu\lambda}+\partial_{\nu}B_{\lambda\mu}+\partial_{\lambda}B_{\mu\nu}.\label{curvature}
\end{eqnarray}
This antisymmetric tensor naturally plays the role of a generalized Faraday (field-strength) tensor in ``2-form electromagnetism"~\cite{Henneaux}. 

A topological invariant can be defined by integrating the 3-form curvature tensor $\mathcal{H}\!=\!\text{d} B$ over a three-dimensional compact manifold, such as the sphere $S^{3}$, 
\begin{eqnarray}\label{DD}
	\mathcal{DD}=\frac{1}{2 \pi^{2}}\int_{S^3} \text{d}^{3}x\, \mathcal{H}_{xyz}.
\end{eqnarray}
This topological invariant, which is known as the Dixmier-Douady ($\mathcal{DD}$) invariant in the literature~\cite{Murray1,Murray2}, is related to the (first) Dixmier-Douady class of U(1) ``bundle gerbes" ~\cite{Waldorf}; see also Refs~\cite{Gawedski,Monaco,Szabo} for applications of bundle gerbes in the context of topological insulators. Similarly to the first Chern number in Eq.~\eqref{Chern_def}, the $\mathcal{DD}$ invariant is an integer, which originates from the third cohomology group, $H^{3}(S^{3})\!=\!\mathbb{Z}$. Finally, we point out that high-order $\mathcal{DD}$ invariants can be obtained by considering higher powers of the curvature $\mathcal{H}$; for instance a ``second $\mathcal{DD}$ invariant" can be obtained by integrating $\mathcal{H}^2\!=\!\mathcal{H}\!\wedge\!\mathcal{H}$ over a 6-dimensional space; see Table~\ref{Table_one}.

\subsection{2-form tensor connections from Bloch states}\label{2formBloch}

In this Section, we generalize the construction of Section~\ref{construction_vector}, in view of building 2-form connections from Bloch states.

Inspired by Ref.~\cite{Pacheva}, we generalize the expression in Eq.~\eqref{vectorB} as
\begin{equation}\label{tensorB}
B_{\mu\nu}=\frac{i}{3}\,\epsilon^{jkl} \phi_{j}^{\aleph}\partial_{\mu}\phi_{k}^{\aleph}\partial_{\nu}\phi_{l}^{\aleph}, \quad j,k,l=\{1,2,3\},
\end{equation}
where we introduced three sets of scalar fields, $\phi_{1,2,3}^{\aleph}$, which all depend on some Bloch states of interest [see Section~\ref{construction_vector}]. We note that the expression in Eq.~\eqref{tensorB} indeed describes a completely antisymmetric tensor $B_{\mu\nu}\!=\!-B_{\nu\mu}$; see Eq.~\eqref{B_anti}.

In this work, we make the specific choice of considering a mixed set of pseudo-real and complex scalar fields, $\phi_{1,2,3}$, and we set $\aleph\!=\!1$ in Eq.~\eqref{tensorB} for simplicity. Specifically, we impose that these three fields satisfy the gauge transformations
\begin{align}\label{gauge7}
&\phi_{1}\equiv \varphi\rightarrow \varphi+\alpha(\bf q), \nonumber \\
&\phi_{2}\rightarrow e^{-i \alpha(\bf q)} \phi_{2}, \qquad \phi_{3}\rightarrow e^{i \alpha(\bf q)} \phi_{3},
\end{align}
where $\phi_{1}\!\equiv\!\varphi$ is a pseudo-real scalar field [Eq.~\eqref{scalar-field}] and where $\phi_{2,3}$ are complex fields. Under these assumptions, one verifies that the field $B_{\mu\nu}$ defined in Eq.~\eqref{tensorB} indeed transforms as a tensor gauge field [Eq.~\eqref{BB-field}] under a gauge transformation.
Here, the quantity $\Lambda_{\mu\nu}$ entering Eq.~\eqref{BB-field} is a non-trivial function of the fields $\phi_{1,2,3}$ and $\alpha(\bf q)$, i.e.~it contains the gauge redundancy of the scalar fields. 

In the following, we will call ``tensor Berry connection", the field $B_{\mu\nu}$ defined in Eq.~\eqref{tensorB}, with the three scalar fields satisfying Eq.~\eqref{gauge7}. We point out that other choices could be made for the scalar fields entering the construction described above; indeed, many more fields $\phi_{j}^{\aleph}$ could be introduced, and different structures could arise depending on their pseudo-real/complex nature. We leave the exploration of these diverse structures as an outlook for future work.

\subsection{A few applications of the tensor Berry connection}\label{Applications} 

We now explore the tensor Berry connection defined in Eq.~\eqref{tensorB}, by studying a series of microscopic models. We will discuss how the introduction of a tensor Berry connection provides a novel view on topological systems in class A and AIII.

\subsubsection{Chern insulators}\label{section:Chern}

We start by considering a two-dimensional model given by the following two-band Hamiltonian
\begin{align}
&H_{2D}=R_1 \sigma_1 + R_2 \sigma_2 + R_3 \sigma_3,\\
&R_{1}=\sin q_{x}, R_{2}=\sin q_{y} , R_{3}=M+\cos q_{x}+\cos q_{y} , \notag
\end{align}
where $\sigma_{1,2,3}$ are three Pauli matrices and $q_{x,y}$ are the quasi-momenta. When the 2-band spectrum is gapped, the model describes a Chern insulator, i.e.~a time-reversal broken topological state (class A).
The eigenstates of the lower band can be written as~\cite{Varma}
\begin{align}
&|u_{-}\rangle=(u_{1},u_{2})^{\top},  \notag \\
&u_{1}=\frac{R_{3}-|E|}{\sqrt{2|E|(|E|-R_{3})}}, \hspace{0.1cm}u_{2}=\frac{R_{1}+i R_{2}}{\sqrt{2|E|(|E|-R_{3})}},\label{u1u2}
\end{align}
and their energies are given by $E\!=\!-\sqrt{R_{1}^{2}+R_{2}^{2}+R_{3}^{2}}$.
The topological invariant characterizing the (Chern) insulating state is the first Chern number [Eq.~\eqref{Chern_def}],
\be
\nu^{1}=\frac{1}{2 \pi} \int_{\text{FBZ}} \text{d}q_x\text{d}q_y\, \mathcal{F}_{xy} ({\bf q}),\label{Chern_qspace}
\ee
where $\mathcal{F}_{xy}$ denotes the (2-form) Berry curvature, and where the integral is performed over the 2D FBZ (i.e.~a torus $\mathbb{T}^2$). In the present case, it can be simply written as 
\begin{equation}
\nu^{1}\!=\!- {\rm sgn} (M), \text{  for $0< |M| < 2$}. \label{Chern_values}
\end{equation}

We now demonstrate that this topological invariant can be reinterpreted as a ``2D Zak phase", associated with a tensor Berry connection $B_{\mu\nu}$. To do so, we first define the three scalar fields entering the definition of the tensor Berry connection in Eq.~\eqref{tensorB} as
\begin{equation}\label{gauge9}
\phi_{1}\equiv\varphi\equiv -i\log u_{1},\hspace{0.3cm}\phi_{2}\equiv u_{2}^{*}, \hspace{0.3cm}
\phi_{3}\equiv u_{2},
\end{equation}
where the components $u_{1,2}$ are defined in Eq.~\eqref{u1u2}.
Since the lower-band eigenstates transform as $|u_{-}\rangle\!\rightarrow\!e^{i \alpha(\bf q)}|u_{-}\rangle$ under a local gauge transformation, the scalar fields $\phi_{1,2,3}$ in Eq.~\eqref{gauge9} indeed satisfy the generalized transformations introduced in Eq.~\eqref{gauge7}.

In direct analogy with the conventional Zak phase in Eq.~\eqref{Zak_def}, one can introduce a ``2D Zak phase" $\gamma^{\text{2D}}$ associated with the tensor Berry connection $B_{\mu\nu}$ by integrating the latter over the 2D first Brillouin zone. First of all, we note that by inserting the Bloch wavefunction defined in Eq~\eqref{u1u2} into the definition of $B_{\mu\nu}$ [Eqs.~\eqref{tensorB} and \eqref{gauge9}], one obtains
\begin{equation}
\frac{3}{2 \pi^{2}}\int_{\text{FBZ}} \text{d} q_x \text{d} q_y \, {\rm Re}\left( B_{xy}\right) = \nu^{1},\label{Chern_B_text}
\end{equation}
where $\nu^{1}$ denotes the first Chern number for the lower band. However, special care is required to define a gauge-independent quantity:~following the analysis presented in Appendix, we define the (gauge-invariant) 2D Zak phase as
\begin{equation}
\gamma^{\text{2D}}=-\frac{3}{2 \pi^{2}}\int_{\text{FBZ}} \text{d} q_x \text{d} q_y \, {\rm Re} \left (  B_{xy}-\varphi \mathcal{F}_{xy} \right)= 2\nu^{1},\label{2DZak_Chern}
\end{equation}
where ${\rm Re} (\cdot)$ denotes the real part of the integrand, and where we introduced the term $-\varphi \mathcal{F}_{xy}$ to ensure gauge invariance. Equations~\eqref{Chern_B_text}-\eqref{2DZak_Chern} indicate that the integral of ${\rm Re}\left( B_{xy}\right)$ is proportional to the Chern number $\nu^{1}$ up to a trivial shift; see Appendix. This result is non-trivial and rather surprising, as it connects two quantities of very different nature:~While the conventional Berry curvature $\mathcal{F}$, which enters the definition of the Chern number [Eq.~\eqref{Chern_qspace}], is gauge invariant and defines a closed differential form ($\text{d} \mathcal{F}\!=\!0$), the tensor Berry connection $B$ that is integrated in Eqs.~\eqref{Chern_B_text}-\eqref{2DZak_Chern} is gauge dependent and its exterior derivative is non-trivial, $\text{d} B\!=\! \mathcal{H}$.

\subsubsection{3D topological insulators in class AIII}

We now consider a model for 3D topological insulators in class AIII, where time-reversal and particle-hole symmetries are both broken, while preserving chiral symmetry~\cite{Ryu,Neupert}.
The Hamiltonian is taken in the form~\cite{Neupert}
\begin{align}\label{3D}
H_{\text{3D}}&=\textbf{d} \cdot \boldsymbol{\lambda}=
d_{1}\lambda_{4}+d_{2}\lambda_{5}+d_{3}\lambda_{6}+d_{4}\lambda_{7} \\
&= \left({\begin{array}{ccc}
	0 & 0 & d_{1}- i d_{2} \\
	0 & 0 & d_{3}- i d_{4} \\
	d_{1}+i d_{2} & d_{3}+ i d_{4} & 0 \\
	\end{array} } \right), \nonumber
\end{align}
where the $\lambda$ matrices are $3\times3$ Gell-Mann matrices~\cite{Gell-Mann}, and where 
\begin{align}
&d_{1}=\sin q_{x} , \quad d_{2}=\sin q_{y} , \quad d_{3}=\sin q_{z} , \notag \\
&d_{4}=M-\cos q_{x}-\cos q_{y}-\cos q_{z}.\label{3Dds}
\end{align}
The corresponding lower-band eigenstates are given by
\begin{align}
&\vert u_{-} \rangle=\frac{1}{\sqrt{2}}(u_{1},u_{2},-1)^{\top}, \label{u3D} \\
&u_{1}=\frac{d_{1}-i d_{2}}{|E|}, \, u_{2}=\frac{d_{3}-i d_{4}}{|E|},\notag
\end{align}
and their energies read $E\!=\!-\sqrt{d_{1}^{2}+d_{2}^{2}+d_{3}^{2}+d_{4}^{2}}$.

The topological invariant characterizing such 3D topological insulators~\cite{Ryu,Neupert} is given by a winding number,
\begin{align}
w^{3}=\frac{1}{12\pi^{2}}\int_{\text{FBZ}}\text{d}^{3}q\,\epsilon^{jklm}\epsilon^{\mu\nu\lambda} \frac{1}{|E|^{4}}\,
d_{j}\partial_{\mu}d_{k}\partial_{\nu}d_{l}\partial_{\lambda}d_{m}.\label{3Dwinding_def}
\end{align}
For the model in Eqs.~\eqref{3D}-\eqref{3Dds}, this yields 
\begin{align}
w^{3}=-1 \text{ for } 1<|M|<3 , \quad w^{3}=2 \text{ for } |M|<3 , 
\end{align}
and zero otherwise; the gap closes at $|M|\!=\!1$ and $|M|\!=\!3$. We emphasize that the winding number $w^{3}$ can be written in terms of the (conventional) Berry connection $A_{\mu}$, through the so-called Chern-Simons form~\cite{Neupert}.

We now demonstrate that this topological invariant is in fact related to the \emph{curvature of a tensor Berry connection}. Following the same approach as for the previous example, we first identify three scalar fields related to the lower-band eigenstates in Eq.~\eqref{u3D},
\begin{equation}\label{gauge8}
\phi_{1}\equiv -i\log u_{2},\hspace{0.3cm}\phi_{2}\equiv u_{1}^{*}, \hspace{0.3cm}
\phi_{3}\equiv u_{1}.
\end{equation}
These fields indeed satisfy the generalized transformations defined in Eq.~\eqref{gauge7} and they provide an explicit expression for the tensor field $B_{\mu\nu}$ in Eq.~\eqref{tensorB}, in terms of the Bloch states components $u_{1,2}$ defined in Eq.~\eqref{u3D}. One can then evaluate the components of the curvature associated with the tensor connection, $\text{d}B\!=\! \mathcal{H}$, which yields
\begin{align}
&\mathcal{H}_{xyz}=\Theta/\Delta^{2}, \\
&\Theta=M\cos q_{x} \cos q_{y} \cos q_{z} \nonumber \\
&\qquad - (\cos q_{x}\cos q_{y}+\cos q_{y}\cos q_{z}+\cos q_{z}\cos q_{x}), \nonumber \\
&\Delta=3+M^{2}-2 M(\cos q_{x}+\cos q_{y}+\cos q_{z}) \nonumber \\
&\qquad +2(\cos q_{x}\cos q_{y}+\cos q_{y}\cos q_{z}+\cos q_{z}\cos q_{x}).\notag
\end{align}
Finally, one evaluates the topological invariant associated with this 3-form curvature, the $\mathcal{DD}$ invariant [Eq.~\eqref{DD}]
\begin{equation}
\mathcal{DD}=\frac{1}{2 \pi^{2}}\int_{\text{FBZ}} \text{d}q_x \text{d}q_y \text{d}q_z \, \mathcal{H}_{xyz}=w^{3},\label{DD_3DTI}
\end{equation}
and we find that it exactly corresponds to the winding number of Eq.~\eqref{3Dwinding_def}. This intriguing result provides an instructive interpretation of the 3D winding number~\cite{Ryu,Neupert}:~this topological number can be seen as the topological charge of a tensor monopole generating a generalized magnetic field $\mathcal{H}\!=\!\text{d} B$ in momentum space; in order words, $w^{3}$ corresponds to the total (quantized) flux of the field $\mathcal{H}\!=\!\text{d} B$ through the 3D Brillouin zone [Eq.~\eqref{DD_3DTI}]. In this sense, the winding number $w^{3}$ constitutes the 3D analogue of the first Chern number, which corresponds to the total flux of the Berry-curvature field $\mathcal{F}$ over a 2D Brillouin zone. Importantly, this analogy only becomes apparent when introducing the tensor Berry connection $B$, and its related curvature $\mathcal{H}\!=\!\text{d} B$, as proposed in this work. 

Before closing this paragraph, we note that a 3D lattice system displaying a uniform $\mathcal{H}$ field would exhibit the analogue of 3D Landau levels in momentum space (in direct analogy with 2D Bloch bands featuring a uniform Berry curvature $\mathcal{F}$; see Ref.~\cite{Price_LLmomentum,Roy,Claassen_LLmomentum,CL_LL}). In this sense, our approach suggests an interesting route for the exploration of 3D Landau levels in topological insulators.

\subsubsection{Tensor monopoles in 4D Weyl semimetals}

A 3D Weyl semimetal is characterized by a nodal (Weyl) point in momentum space, which can be interpreted as a source of Berry curvature $\mathcal{F}$:~the Weyl point acts as a fictitious (Dirac) monopole in 3D momentum space~\cite{Turner,Armitage}. Similarly, as we recently proposed in Ref.~\cite{Palumbo-Goldman}, a Weyl node defined over a 4D momentum space can be the source of a 3-form curvature $\mathcal{H}\!=\!\text{d} B$, associated with a tensor gauge field $B$. The resulting ``4D tensor monopole", which was originally studied in high-energy physics~\cite{Nepomechie,Teitelboim}, is characterized by a topological charge given by the $\mathcal{DD}$ invariant in Eq.~\eqref{DD}; see Ref.~\cite{Palumbo-Goldman}. It is the aim of this paragraph to clarify the link between such a 4D tensor monopole and the tensor Berry connection of a 4D quantum system [Eq.~\eqref{tensorB}].

Let us first introduce a 4D Weyl-type Hamiltonian~\cite{Palumbo-Goldman}
\begin{align}\label{4D}
H_{\text{4D}}&=q_{x}\lambda_{1}+q_{y}\lambda_{2}+q_{z}\lambda_{6}+q_{w}\lambda_{7}^{*} \\
&= \left({\begin{array}{ccc}
	0 & q_{x}- i q_{y} & 0 \\
	q_{x}+ i q_{y} & 0 & q_{z}+ i q_{w} \\
	0 & q_{z}- i q_{w} & 0 \\
	\end{array} } \right), \nonumber
\end{align}
where the $\lambda$ matrices are $3\times3$ Gell-Mann matrices~\cite{Gell-Mann} and where $q_{x,y,z,w}$ denote the momenta in 4D space.
The lower-band eigenstates are given by
\begin{align}
&\vert u_{-} \rangle=\frac{1}{\sqrt{2}}(u_{1},-1,u_{2})^{\top},\label{u4D} \\
& u_{1}=\frac{q_{x}-i q_{y}}{|E|}, \, u_{2}=\frac{q_{z}-i q_{w}}{|E|},\notag
\end{align}
and their dispersion reads $E\!=\!-\sqrt{q_{x}^{2}+q_{y}^{2}+q_{z}^{2}+q_{w}^{2}}$.

Following the same approach as for the previous examples, we build a tensor connection $B_{\mu\nu}$ through Eq.~\eqref{tensorB}, and we use the same definition for the scalar fields as in Eq.~\eqref{gauge8}, where $u_{1,2}$ now refer to the components of the state in Eq.~\eqref{u4D}. There are a few representative components of the resulting tensor Berry connection:
\begin{eqnarray}
B_{xy}=-\frac{i\left(q_{x}^{2}+q_{y}^{2}+(q_{z}^{2}+q_{w}^{2})\log\Delta\right)}{3(q_{x}^{2}+q_{y}^{2}+q_{z}^{2}+q_{w}^{2})^{2}}, \hspace{2.7cm} \\
B_{yz}=\frac{q_{x}\left(q_{x}^{2}+q_{y}^{2}+q_{w}^{2}+i q_{z}q_{w}+q_{z}(-i q_{w}+q_{z})\log\Delta\right)}{3(q_{w}+i q_{z})(q_{x}^{2}+q_{y}^{2}+q_{z}^{2}+q_{w}^{2})^{2}}, \nonumber \\
B_{zx}=\frac{q_{y}\left(q_{x}^{2}+q_{y}^{2}+q_{w}^{2}+i q_{z}q_{w}+q_{z}(-i q_{w}+q_{z})\log\Delta\right)}{3(q_{w}+i q_{z})(q_{x}^{2}+q_{y}^{2}+q_{z}^{2}+q_{w}^{2})^{2}},\nonumber
\end{eqnarray}
where
\begin{eqnarray}
\Delta= \frac{-i q_{w}+q_{z}}{\sqrt{2(q_{x}^{2}+q_{y}^{2}+q_{z}^{2}+q_{w}^{2})}}.
\end{eqnarray}

After a tedious calculation, one obtains a simple expression  for the related 3-form curvature $\mathcal{H}\!=\!\text{d} B$,
\begin{align}
\mathcal{H}_{\mu\nu\lambda}&=\partial_{\mu}B_{\nu\lambda}+\partial_{\nu}B_{\lambda\mu}+\partial_{\lambda}B_{\mu\nu} \nonumber \\
&=\epsilon_{\mu\nu\lambda\gamma}\,\frac{q_{\gamma}}{(q_{x}^{2}+q_{y}^{2}+q_{z}^{2}+q_{w}^{2})^{2}}.\label{Hmonopole}
\end{align}
As announced above, the $\mathcal{H}$ field in Eq.~\eqref{Hmonopole} indeed acts as a generalized magnetic field in momentum space, which is generated by a fictitious monopole located at the origin (i.e.~at the 4D Weyl node). In particular, integrating this $\mathcal{H}$ field over a sphere $S^3$ that surrounds the Weyl node yields the topological charge of this tensor monopole; this is the $\mathcal{DD}$ invariant associated with the $\mathcal{H}$ field [Eq.~\eqref{DD}], which, as we mentioned earlier, is related to the third cohomology group of the 3-sphere,~$H^{3}(S^{3})\!=\!\mathbb{Z}$.

In the context of 3D Weyl semimetals, it is well known that taking a slice close to a Weyl point ($q_z\!=\!c$ constant) yields a gapped 2D Chern insulator~\cite{Turner,Armitage}. Now, consider a 4D lattice system exhibiting a single nodal point in momentum space, which is locally described by the Hamiltonian in Eq.~\eqref{4D}. Then, taking a 3D section close to this ``4D Weyl node", by setting $q_{w}\!=\!{\rm c}$ constant, would lead to a gapped 3D topological insulator in class AIII; this is due to the fact that the system in Eq.~\eqref{4D} preserves chiral symmetry~\cite{Palumbo-Goldman}.  

It is interesting to extend this discussion to infinite parameter spaces, i.e.~by assuming that the momenta $q_{\mu}$ in Eq.~\eqref{4D} are not restricted to a Brillouin zone. First, let us recall that a gapped nodal point in 2D space describes a massive Dirac fermion of mass $M$, whose topological nature is captured by a semi-integer invariant 
\begin{equation}
\nu^1\!=\frac{1}{2 \pi} \int_{- \infty}^{\infty} \text{d} q_x \text{d}q_y\, \mathcal{F}_{xy}=(1/2)\, {\rm sgn}\,M ,
\end{equation}
where $\mathcal{F}_{xy}$ is the Berry curvature related to the lower band. Similarly, we have verified that taking a 3D section of the 4D model defined in Eq.~\eqref{4D} would result in a massive 3D Weyl fermion, whose topological nature is captured by the semi-integer $\mathcal{DD}$ invariant  
\begin{equation}
\mathcal{DD}=\frac{1}{2 \pi^{2}}\int_{- \infty}^{\infty} \text{d}q_x \text{d}q_y \text{d}q_z \, \mathcal{H}_{xyz}=(1/2)\, {\rm sgn}\,c,\label{DDhalf}
\end{equation}
for a section $q_w\!=\!c$ constant. In the same manner that a single massive 2D Dirac fermion lives on the gapped boundary of a 3D topological insulator~\cite{Qi}, the massive 3D Weyl fermion characterized by Eq.~\eqref{DDhalf} could be found at the gapped boundary of a 4D topological insulator.

Finally, we point out that the 4D tensor monopole associated with the 2-form Berry connection $B_{\mu\nu}$ can be generalized to higher dimensions, by introducing higher-order ($p$-form) Berry connections; see the next Section~\ref{higher}.



\subsection{Extension to higher-order tensor gauge fields} \label{higher}

In this section, we briefly describe possible generalizations to higher-order tensor Berry connections. 

Higher-order tensor gauge fields have been introduced in quantum field theory and string theory~\cite{Montero,Gaiotto}, as a natural generalization to $2$-form tensor connections~\cite{Henneaux}. The simplest higher-order tensor field corresponds to the $3$-form connection, denoted $C$, whose components transform as
\begin{eqnarray}\label{CC-field}
C_{\mu\nu\lambda}\rightarrow C_{\mu\nu\lambda}+\Lambda_{\mu\nu\lambda},
\end{eqnarray}
under an Abelian gauge transformation; here $\Lambda_{\mu\nu\lambda}$ is an arbitrary 3-form field. In specific cases, $\Lambda_{\mu\nu\lambda}$ can itself be expressed as an exterior derivative, namely $\Lambda_{\mu\nu\lambda}\!=\partial_{\mu}\zeta_{\nu\lambda}+\partial_{\nu}\zeta_{\lambda\mu}+\partial_{\lambda}\zeta_{\mu\nu}$, where $\zeta_{\mu\nu}$ denotes the components of a 2-form.

Generalizing the notion of Wilson surfaces [Eq.~\eqref{Wilsonsurface}], one can define a ``Wilson volume" associated with the $C$ field,
\begin{eqnarray}
W_{V}=\exp\left(\iiint \text{d}x^{\mu}\wedge \text{d}x^{\nu}\wedge \text{d}x^{\lambda}\, C_{\mu\nu\lambda}\right).
\end{eqnarray}
This generalized geometric phase is related to a higher-order gerbe holonomy~\cite{Murray3}. Using Stokes' theorem, one can express the Wilson volume in terms of a 4-form curvature $\mathcal{V}\!=\!\text{d}C$, whose components are given by
\begin{eqnarray}
\mathcal{V}_{\mu\nu\lambda\delta}=\partial_{\mu}C_{\nu\lambda\delta}-\partial_{\nu}C_{\lambda\delta\mu}+\partial_{\lambda}C_{\delta\mu\nu}-\partial_{\delta}C_{\mu\nu\lambda}.
\end{eqnarray}
The integral of this 4-form curvature over a compact 4D manifold defines a topological invariant, which naturally generalizes the $\mathcal{DD}$ invariant [Eq.~\eqref{DD}].

Following the same approach as in Section~\ref{2formBloch}, we write the components of the higher-order tensor $C$ in terms of a set of scalar fields
\begin{eqnarray}\label{highertensorC}
C_{\mu\nu\lambda}=\frac{i}{4}\,\epsilon^{jklm} \phi_{j}^{\aleph}\partial_{\mu}\phi_{k}^{\aleph}\partial_{\nu}\phi_{l}^{\aleph}\partial_{\lambda}\phi_{m}^{\aleph}, 
\end{eqnarray}
where $j,k,l,m\!=\!\{1,2,3,4\}$; the resulting tensor is completely antisymmetric, similarly to the tensor gauge field $B_{\mu\nu}$. As for the 2-form Berry connection of Eq.~\eqref{tensorB}, various choices can be envisaged for the scalar fields. For instance, one could consider a set of complex scalar fields satisfying the following gauge transformations,
\begin{align}\label{gauge5}
&\phi_{1}^{\aleph}\rightarrow e^{i \alpha(\bf q)}\phi_{1}^{\aleph}, \qquad \phi_{2}^{\aleph}\rightarrow e^{i \beta(\bf q)} \phi_{2}^{\aleph}, \nonumber \\
&\phi_{3}^{\aleph}\rightarrow e^{i \gamma(\bf q)} \phi_{3}^{\aleph}, \qquad
\phi_{4}^{\aleph}\rightarrow e^{i \delta(\bf q)} \phi_{4}^{\aleph},
\end{align}
upon imposing the condition $\alpha({\bf q})+\beta({\bf q})+\gamma({\bf q})+\delta({\bf q})\!=\!0$.  One could also consider a mixture of complex and pseudo-real scalar fields, satisfying the gauge transformations
\begin{align}\label{gauge10}
&\phi_{1}^{\aleph}\equiv \varphi_{1}^{\aleph}\rightarrow \varphi_{1}^{\aleph}+\alpha({\bf q}), \qquad \phi_{2}^{\aleph}\equiv \varphi_{2}^{\aleph}\rightarrow \varphi_{2}^{\aleph}+\alpha({\bf q}), \nonumber \\ 
&\phi_{3}^{\aleph}\rightarrow e^{-i \alpha(\bf q)} \phi_{3}^{\aleph}, \qquad \qquad \quad
\phi_{4}^{\aleph}\rightarrow e^{i \alpha(\bf q)} \phi_{4}^{\aleph}. 
\end{align}
As in Section~\ref{Applications}, these scalar fields could then be related to the components of Bloch states of interest, in order to define their corresponding ``3-form Berry connection".

The introduction of a 3-form Berry connection allows one to define a ``3D Zak phase",  
\begin{equation}
\gamma^{\text{3D}}\sim\int_{\text{FBZ}} \text{d}q_x \text{d}q_y \text{d}q_z\, {\rm Re}\, \left (C_{xyz} \right ),\label{3DZak}
\end{equation}
where the integration is performed over a three-dimensional first Brillouin zone of a quantum system of interest.
Inspired by the results of Section~\ref{Applications}, we suggest that this quantity could be related to the topological invariants of 3D lattice systems.

Introducing the curvature of the 3-form Berry connection, $\mathcal{V}\!=\!\text{d}C$, one can also define a topological invariant for 4D lattice systems through the integral
\begin{equation}
\int_{\text{FBZ}} \text{d}q_x \text{d}q_y \text{d}q_z \text{d}q_w \, \mathcal{V}_{xyzw},
\end{equation}
which is associated with a higher-order gerbe bundle over quasi-momentum space. In fact, this invariant was shown to be directly proportional to the second Chern number $\nu^{2}$ of a 5D tensor monopole~\cite{Yang,Hasebe}. This suggests that higher-order tensor Berry connections could represent suitable quantities for the exploration and characterization of higher-dimensional topological states, with possible applications in quantum-engineered setups~\cite{Goldman2,Goldman3,CHLee6D,Price6D}.

\section{Tensor connections and their topological invariants}\label{Section_table}

In this last Section, we provide a general overview of the zoo of topological invariants that could emerge from the existence of tensor Berry connections.

As we briefly reviewed in Section~\ref{section_vector}, the conventional (vector) Berry connection is associated with a variety of topological invariants, whose relevance depend on the dimensionality (and symmetries) of the underlying model~\cite{Xiao,Qi,Ryuclass,Ryu}. Examples of such invariants are provided in the first row of Table~\ref{Table_one}:~the line integral of the (vector) Berry connection $A$ over the first Brillouin zone (FBZ) of a 1D system is known as the Zak phase~\cite{Zak,Xiao}; the surface integral of the curvature $\mathcal{F}\!=\!\text{d}A$ over the FBZ of a 2D system is known as the first Chern number~\cite{TKNN,Qi}; the integral of the curvature squared  $\mathcal{F}^2$ over a 4D FBZ is known as the second Chern number~\cite{Zhang4D,Goldman3}; the integral of  $\mathcal{F}^3$ over a 6D FBZ is known as the third Chern number~\cite{Zhang4D,Goldman3}; so on and so forth. We note that other types of invariants can be constructed from the conventional Berry connection, such as the Chern-Simons invariants~\cite{Zhang,Neupert} in odd-dimensional systems.

The existence of a (2-form) tensor Berry connection $B_{\mu \nu}$ leads to a new series of invariants; see second row of Table~\ref{Table_one}. As previously discussed, the 2D Zak phase [Eq.~\eqref{2DZak_Chern}] can be defined as the surface integral of $B$ over a 2D FBZ; in this work, we have presented an example where this 2D Zak phase is strictly equivalent to the 1st Chern number associated with the conventional (vector) Berry connection [Eq.~\eqref{2DZak_Chern}]. The curvature of the tensor Berry connection, $\mathcal{H}\!=\!\text{d}B$, gives rise to the $\mathcal{D}\mathcal{D}$ invariant upon integration over a 3D FBZ; we have shown how this invariant connects to the winding number characterizing 3D topological insulators in class AIII [Eq.~\eqref{DD_3DTI}]. One can then consider the curvature squared, $\mathcal{H}^2$, which constitutes a 6-form; the integral of this quantity over a 6D FBZ should give rise to a topological invariant characterizing 6D topological systems; we postulate that the resulting invariant can potentially be related to the 3rd Chern number (first line of Table~\ref{Table_one}). Considering $\mathcal{H}^3$, one obtains a 9-form, which could be used to characterize the topology of 9D models.

Section~\ref{higher} introduced the 3-form connection $C$, whose 3D Zak phase [Eq.~\eqref{3DZak}] is defined over a 3D FBZ; see third row of Table~\ref{Table_one}. The curvature of $C$, $\mathcal{V}\!=\!\text{d}C$, constitutes a 4-form whose integration over a 4D FBZ is reminiscent of the 2nd Chern number~\cite{Hasebe}. Taking the squared of the curvature $\mathcal{V}^2$ yields a 8-form, whose integration over a 8D FBZ could potentially be related to the 4th Chern number. Similarly, integrating the 12-form $\mathcal{V}^3$ over a 12-dimensional FBZ could lead to an invariant reminiscent of the 6th Chern number.

\begin{table}[!ht]
	\renewcommand{\arraystretch}{2}
	\begin{tabular}{|c|c|c|c|c|}
		\hline
		$\mathbb{Z}$  & \multicolumn{1}{l|}{$\int \mathfrak{a}$} & \multicolumn{1}{l|}{$\int \text{d}\mathfrak{a}$} & \multicolumn{1}{l|}{$\int \text{d}\mathfrak{a}\!\wedge\!\text{d}\mathfrak{a}$} & \multicolumn{1}{l|}{$\int \text{d}\mathfrak{a}\!\wedge \text{d}\mathfrak{a}\!\wedge \text{d}\mathfrak{a}$} \\ \hline
		$A_{\mu}$           & 1 (Zak)                              & 2  (1$^{\text{st}}$ Chern)                            & 4 (2$^{\text{nd}}$ Chern)                                                & 6 (3$^{\text{rd}}$ Chern)                                                 \\ \hline
		$B_{\mu\nu}$        & 2  (2d Zak)                              & 3 ($\mathcal{DD}$)                             & 6 (2$^{\text{nd}}$ $\mathcal{DD}$)                                                & 9 (3$^{\text{rd}}$ $\mathcal{DD}$)                                                 \\ \hline
		$C_{\mu\nu\lambda}$ & 3 (3d Zak)                                & 4                              & 8                                                 & 12                                                \\ \hline
	\end{tabular}
	\renewcommand{\arraystretch}{1}
	\caption{Classification of topological invariants, used to characterize topological matter in classes A and AIII, depending on the dimensionality of the underlying system. Each entry is an integer corresponding to the dimensions of the corresponding first Brillouin zone (FBZ). Invariants are expressed as an integral of a connection $\mathfrak{a}$, or as an integral of its related curvature $\text{d}\mathfrak{a}$, or as an integral of a power $(\text{d}\mathfrak{a})^n\!=\!\text{d}\mathfrak{a} \wedge \cdots \wedge \text{d}\mathfrak{a}$, over the FBZ. The first row displays the topological invariants associated with a vector Berry connection $A_{\mu}$:~the Zak phase in 1D, and the Chern numbers in 2D, 4D and 6D. The second row displays the invariants associated with a (2-form) tensor Berry connection $B_{\mu\nu}$:~the 2D Zak phase in 2D, and the $\mathcal{DD}$ invariants in 3D, 6D and 9D. The third row presents the invariants associated with the higher-order tensor connection $C_{\mu\nu\lambda}$. In some cases, the topology of a given quantum system can be expressed in terms of several (equivalent) invariants:~e.g.~the 2D Zak phase $\int B$ (second row) can be related to the 1st Chern number $\int \text{d}A$ (first row); the invariant $\int \text{d}C$ (third row) is related to the second Chern number $\int \text{d}A\wedge\text{d}A$ (first row); see text. }
%
%
%
\label{Table_one}
\end{table}

\section{Concluding remarks}\label{Section_conclusions}

This work proposed tensorial generalizations of the Berry connection based on a formal approach, which allows for the construction of a wide variety of effective gauge potentials in quantum matter. This includes scalar, vector and ($p$-form) tensor gauge fields, whose relevance depend on the dimensionality of the underlying momentum (or parameter) space. While these different connections are gauge dependent, one has presented examples of gauge-invariant and topological quantities that can be naturally derived from them. Interestingly, this approach leads to a novel interpretation of well-known topological invariants, such as the winding number of 1D topological insulators (in terms of a generalized Zak phase), the Chern number (in terms of a 2D Zak phase), and the winding number of 3D topological insulators (in terms of a $\mathcal{DD}$ invariant). Moreover, the study of tensor Berry connections allows for the identification of exotic topological defects in higher-dimensional quantum systems, such as the 4D monopole of Ref.~\cite{Palumbo-Goldman}, which envisages novel forms of topological semimetals in 4D lattice systems, as well as intriguing gapped phases in 3D lattices characterized by the $\mathcal{DD}$ invariant. Of particular interest is the exploration of isolated 3D Weyl-type fermions, at the gapped boundaries of 4D lattice systems, whose topology is characterized by the $\mathcal{DD}$ invariant; direct extensions to 5D systems are also possible and appealing. We point out that such higher-dimensional lattice systems could be engineered in cold-atom setups~\cite{Goldman3,CHLee6D,Price6D,Price4Dnew} and photonics\cite{Ozawa4D,Ozawa_review}, using the concept of synthetic dimensions~\cite{Goldman2}. 

More generally, our formalism sheds new light on the relation between (gapped) topological insulators and topological semimetals:~In the same way that a 2D Chern insulator shares the same topological nature as a 3D Weyl semimetal, for each $d$-dimensional gapped topological state in class A and AIII (with $d\!>\!1$), one can associate a topological semimetal in $(d+1)$-dimensions; the latter semimetal typically exhibits a tensor monopole, namely, a source of a tensor (or higher-order tensor) Berry connection.

Finally, our work builds a novel bridge between the condensed-matter and high-energy communities, by demonstrating how momentum-space Kalb-Ramond fields naturally emerge in non-relativistic quantum physics. This novel connection suggests intriguing directions of research in the realm of ultra quantum matter.

 {\bf Acknowledgments: }
The authors are pleased to acknowledge discussions with Alexander Altland, Andrea Cappelli, Sebastian Diehl, Hannah M. Price and Martin Zirnbauer. They also thank the journal's Referee for their insightful remarks on their manuscript, in particular, regarding the gauge invariance of higher-form Zak phases. The authors are supported by the ERC Starting Grant TopoCold and the Fonds De La Recherche Scientifique (FRS-FNRS) (Belgium).

\section*{Appendix}

In this Appendix, we provide more details regarding the tensor Berry connection of two-band Chern insulators [Section~\ref{section:Chern}], and on the gauge invariance of the related Zak phase.

First of all, we note that by inserting the Bloch wavefunction defined in Eq~\eqref{u1u2} into the definition of $B_{\mu\nu}$ [Eqs.~\eqref{tensorB} and \eqref{gauge9}], one obtains
\begin{equation}
\frac{3}{2 \pi^{2}}\int_{\text{FBZ}} \text{d} q_x \text{d} q_y \, {\rm Re}\left( B_{xy}\right) = \nu^{1},\label{Chern_B}
\end{equation}
where $\nu^{1}$ denotes the first Chern number for the lower band. However care is required when exploiting this relation, since the left-hand-side of Eq.~\eqref{Chern_B} is in fact gauge dependent, as we now discuss.
To see this, we determine the explicit form of the shift $\Lambda_{\mu\nu}$ entering Eq.~\eqref{BB-field}, in terms of the scalar fields entering the definition of the tensor Berry connection. After several algebraic manipulations, we obtain
\begin{align}
\Lambda_{\mu\nu}=\partial_{\mu}\left[\frac{1}{6}\alpha^{2}\partial_{\nu}(\phi_{2}\phi_{3})+\frac{i}{3}\alpha \phi_{3}\partial_{\nu}\phi_{2}- \frac{i}{3}\alpha \phi_{2}\partial_{\nu}\phi_{3}\nonumber \right. \\
\left.
+\frac{1}{3}\alpha \phi_{1}\partial_{\nu}(\phi_{2}\phi_{3})-\frac{2}{3}\alpha\phi_{2}\phi_{3}\partial_{\nu}\phi_{1}\right]-
\nonumber  \\
\partial_{\nu}\left[\frac{1}{6}\alpha^{2}\partial_{\mu}(\phi_{2}\phi_{3})+\frac{i}{3}\alpha \phi_{3}\partial_{\mu}\phi_{2}- \frac{i}{3}\alpha \phi_{2}\partial_{\mu}\phi_{3}
\nonumber \right. \\
\left.
+\frac{1}{3}\alpha \phi_{1}\partial_{\mu}(\phi_{2}\phi_{3})-\frac{2}{3}\alpha\phi_{2}\phi_{3}\partial_{\mu}\phi_{1}\right]+\nonumber \\
\alpha \left[\mathcal{F}_{\mu\nu}+\partial_{\mu}\phi_{1}\partial_{\nu}(\phi_{2}\phi_{3})-\partial_{\nu}\phi_{1}\partial_{\mu}(\phi_{2}\phi_{3})\right],
\end{align}
where $\mathcal{F}_{\mu\nu}$ is the standard (2-form) Berry curvature associated with the Bloch states.

We are ultimately interested in the real part of the shift, since
\begin{eqnarray}
{\rm Re} (B_{\mu\nu})\rightarrow {\rm Re} (B_{\mu\nu})+{\rm Re}(\Lambda_{\mu\nu}),
\end{eqnarray}
and we remind that ${\rm Re} (B_{\mu\nu})$ enters our definition of the 2D Zak phase. In order to simplify the expression for ${\rm Re}(\Lambda_{\mu\nu})$, we note that
\begin{eqnarray}
\phi_{1}\equiv\varphi=-i\log u_{1}=-i\log |u_{1}|+\pi,
\end{eqnarray}
since $u_{1}\!=\!\frac{R_{3}-|E|}{\sqrt{2|E|(|E|-R_{3})}}$ $\left( E\!=\!\sqrt{R_{3}^{2}+R_{2}^{2}+R_{1}^{2}}\right)$ is a wave function that always takes negative real values; this latter result implies that
\begin{eqnarray}
\partial_{\mu}\phi_{1}= -\frac{i}{|u_{1}|}\partial_{\mu} |u_{1}|,
\end{eqnarray}
is always purely imaginary. Moreover, we note that the product $\phi_{2}\phi_{3}\!=\!|\phi_{3}|^{2}$ is always real since $\phi_{2}\!=\!\phi_{3}^{*}$ ($\alpha$ is a real-valued phase). We eventually obtain
\begin{eqnarray}\label{Lambda}
{\rm Re}(\Lambda_{\mu\nu})=\partial_{\mu}\left[\frac{1}{6}\alpha^{2}\partial_{\nu}|\phi_{3}|^{2}+\frac{2}{3}\alpha {\rm Im}(\phi_{3})\partial_{\nu}{\rm Re}(\phi_{3})- \right. \nonumber  \\ 
\left.
\frac{2}{3} \alpha {\rm Re}(\phi_{3})\partial_{\nu}{\rm Im}(\phi_{3})+\frac{\pi}{3}\alpha \partial_{\nu}|\phi_{3}|^{2}\right]-\nonumber \\
\partial_{\nu}\left[\frac{1}{6}\alpha^{2}\partial_{\mu}|\phi_{3}|^{2}+\frac{2}{3}\alpha {\rm Im}(\phi_{3})\partial_{\mu}{\rm Re}(\phi_{3})-\right. \nonumber  \\ 
\left.
\frac{2}{3} \alpha {\rm Re}(\phi_{3})\partial_{\mu}{\rm Im}(\phi_{3})+\frac{\pi}{3}\alpha \partial_{\mu}|\phi_{3}|^{2}\right]+\nonumber \\
{\rm Re}(\alpha \mathcal{F}_{\mu\nu}).\hspace{4.4cm}
\end{eqnarray}
By integrating this quantity over the first Brillouin zone (a compact two-dimensional manifold), we obtain a non-vanishing contribution from the shift,
\begin{equation}
\int_{\text{FBZ}} \text{d} q_x \text{d} q_y  {\rm Re}(\Lambda_{xy})=\int_{\text{FBZ}} \text{d} q_x \text{d} q_y  {\rm Re}(\alpha \mathcal{F}_{xy}).
\end{equation}
Here, we used the fact that the total-derivative terms in Eq.~\eqref{Lambda} disappear upon integration over the closed manifold.

These result allow us to define a gauge-invariant 2D Zak phase. Let us now consider the product $\varphi \mathcal{F}_{\mu\nu}$, which under U(1) transforms as follows
\begin{equation}
\varphi \mathcal{F}_{\mu\nu}\rightarrow \varphi \mathcal{F}_{\mu\nu}+\alpha \mathcal{F}_{\mu\nu},
\end{equation}
such that
\begin{equation}
{\rm Re} (\varphi \mathcal{F}_{\mu\nu})\rightarrow {\rm Re}(\varphi \mathcal{F}_{\mu\nu})+{\rm Re}(\alpha \mathcal{F}_{\mu\nu}).
\end{equation}
This eventually allows us to define the (gauge-invariant) 2D Zak phase
\begin{equation}\label{2DZak}
\gamma^{\text{2D}}=-\frac{3}{2 \pi^{2}}\int_{\text{FBZ}} \text{d} q_x \text{d} q_y \, {\rm Re} \left (  B_{xy}-\varphi \mathcal{F}_{xy} \right)= 2\nu^{1},
\end{equation}
which is completely gauge invariant and is proportional to the first Chern number $\nu^{1}$. Equations~\eqref{Chern_B}-\eqref{2DZak} indicate that the integral of ${\rm Re}\left( B_{xy}\right)$ is proportional to the Chern number $\nu^{1}$ up to a trivial shift.


\end{document}